\documentstyle[twoside,fleqn,npb,epsfig]{article}
\newcommand{\AmS}{{\protect\the\textfont2
  A\kern-.1667em\lower.5ex\hbox{M}\kern-.125emS}}

\title{Gluon topology and the spin structure of the constituent quark}
\author{Steven D. Bass 
\address{Physik Department, Technische Universit\"at M\"unchen, \\
D-85747 Garching, Germany}
        \thanks{This research is supported by BMBF.}
}

\begin{document}

\begin{abstract}
Gluon topology makes a potentially important contribution to the
spin of the constituent quark.

\end{abstract}

\maketitle

\newpage

\section{INTRODUCTION}

The small value of the flavour-singlet axial charge $g_A^{(0)}$
which is extracted from the first moment of $g_1$
(the nucleon's first spin dependent structure function)
\begin{equation}
\left. g^{(0)}_A \right|_{\rm pDIS} = 0.2 - 0.35. 
\end{equation}
has inspired much theoretical and experimental effort to
understand the internal spin structure of the nucleon
\cite{windmolders}.
A key issue \cite{bass99} is the role of the axial anomaly 
in the transition from parton to constituent quark degrees 
of freedom in low energy QCD.
In this paper I explain why some fraction of proton's spin may
be carried by gluon topology.  The topological contribution has 
support only at Bjorken $x$ equal to zero.

In deep inelastic processes the internal structure of the nucleon is
described by the QCD parton model \cite{partal}.
The deep inelastic structure functions may be written as the sum over 
the convolution of ``soft'' quark and gluon parton distributions with 
``hard'' photon-parton scattering coefficients.
The (target dependent) parton distributions describe a flux of quark 
and gluon partons carrying some fraction
$x = p_{+ \rm parton} / p_{+ \rm proton}$
of the proton's momentum
into the hard (target independent) photon-parton interaction which is 
described by the hard scattering coefficients.

In low energy processes the nucleon behaves like a colour neutral system 
of three massive constituent quark quasi-particles interacting 
self consistently with a cloud of virtual pions which is induced by
spontaneous chiral symmetry breaking \cite{cloudy,njlb}.

One of the most challenging problems in particle physics is to understand 
the transition between the fundamental QCD ``current'' quarks and gluons
and the constituent quarks of low-energy QCD.
The fundamental building blocks are the local QCD quark and gluon 
fields together with the non-local structure \cite{Callan} associated with 
gluon topology \cite{rjc}.

Relativistic constituent-quark pion coupling models predict
$g_A^{(0)} \simeq 0.6$ --- two standard deviations greater 
than the value of $g_A^{(0)}|_{\rm pDIS}$ in Eq.(1).
Can we reconcile these two values of $g_A^{(0)}$ without 
abandoning the constituent quark picture of the nucleon ?

\section{GLUON TOPOLOGY AND $g_A^{(0)}$}

The flavour-singlet axial charge $g_A^{(0)}$ is measured by 
the proton forward matrix element of the gauge invariantly
renormalised axial-vector current
\begin{equation}
J^{GI}_{\mu5} = \left(\bar{u}\gamma_\mu\gamma_5u
                  + \bar{d}\gamma_\mu\gamma_5d
                  + \bar{s}\gamma_\mu\gamma_5s\right)_{GI}
\end{equation} 
viz.
\begin{equation}
2m s_\mu g_A^{(0)} = 
\langle p,s|
\ J^{GI}_{\mu5} \ |p,s\rangle _c
\end{equation}
In QCD the axial anomaly \cite{adler,bell} induces various gluonic
contributions to $g_A^{(0)}$.
The flavour-singlet axial-vector current satisfies the anomalous
divergence equation
\begin{equation}
\partial^\mu J^{GI}_{\mu5}
= 2f\partial^\mu K_\mu + \sum_{i=1}^{f} 2im_i \bar{q}_i\gamma_5 q_i
\end{equation}
where
\begin{equation}
K_{\mu} = {g^2 \over 16 \pi^2}
\epsilon_{\mu \nu \rho \sigma}
\biggl[ A^{\nu}_a \biggl( \partial^{\rho} A^{\sigma}_a 
- {1 \over 3} g 
f_{abc} A^{\rho}_b A^{\sigma}_c \biggr) \biggr]
\end{equation}
is a renormalised version of the Chern-Simons current
and $f=3$ is the number of light-flavours.
Eq.(4) allows us to write
\begin{equation}
J_{\mu 5}^{GI} = J_{\mu 5}^{\rm con} + 2f K_{\mu}
\end{equation}
where 
\begin{equation}
\partial^{\mu} K_{\mu} 
= {g^2 \over 32 \pi^2} G_{\mu \nu} {\tilde G}^{\mu \nu}.
\end{equation}
and
\begin{equation}
\partial^\mu J^{\rm con}_{\mu5}
= \sum_{l=1}^{f} 2im_i \bar{q}_l\gamma_5 q_l
\end{equation}
The partially conserved axial-vector current 
$J_{\mu 5}^{\rm con}$ and the Chern-Simons current $K_{\mu}$
are separately gauge dependent.
Gauge transformations shuffle a scale invariant operator 
quantity between the two operators $J_{\mu 5}^{\rm con}$ 
and $K_{\mu}$ whilst keeping $J_{\mu 5}^{GI}$ invariant.

One would like to isolate the gluonic contribution to $g_A^{(0)}$
associated with $K_{\mu}$ and thus write $g_A^{(0)}$ as the sum
of ``quark'' and ``gluonic'' contributions.
Here we have to be careful because of the gauge dependence of $K_{\mu}$.

Whilst $K_{\mu}$ is a gauge dependent operator, its forward 
matrix elements are invariant under the ``small'' gauge 
transformations of perturbative QCD.
In the QCD parton model one finds \cite{efremov,ar,ccm}
\begin{equation}
g_A^{(0)}|_{\rm partons} 
= 
\Biggl(
\sum_q \Delta q - f {\alpha_s \over 2 \pi} \Delta g \Biggr)_{\rm partons}
\end{equation}
Here ${1 \over 2} \Delta q$ and $\Delta g$ are the amount of spin carried 
by quark and gluon partons in the polarised proton.

The polarised gluon contribution to Eq.(9) is characterised by 
the contribution to the first moment of $g_1$ 
from two-quark-jet events carrying large transverse momentum squared
$k_T^2 \sim Q^2$ which are generated by photon-gluon fusion \cite{ccm}.
The polarised quark contribution $\Delta q_{\rm parton}$ 
is associated with the first moment of the measured $g_1$ 
after these two-quark-jet events are subtracted from the total data set.

The QCD parton model formula (9) is not the whole story.  
Choose a covariant gauge. 
When we go beyond perturbation theory, the forward matrix elements 
of $K_{\mu}$ are not invariant under ``large'' gauge transformations 
which change the topological winding number \cite{jaffem}.
The topological winding number is a non-local property of QCD.
It is determined by the gluonic boundary conditions at 
``infinity'' \cite{rjc}
---
 a large surface with boundary which is spacelike with respect 
 to the positions $z_k$ of any operators or fields in the physical
 problem ---
and is insensitive to any local deformations of the gluon field 
$A_{\mu}(z)$ or of the gauge transformation $U(z)$
--- that is, perturbative QCD degrees of freedom.
When we take the Fourier transform to momentum space the topological 
structure induces a light-cone zero-mode which has support only at $x=0$
\cite{bass98}. 
Hence, we are led to consider the possibility that there may be a 
term in $g_1$ which is proportional to $\delta(x)$.

It remains an open question whether the net non-perturbative quantity 
which is shuffled between the $J_{\mu 5}^{\rm con}$ and $K_{\mu}$ 
contributions to $g_A^{(0)}$
under ``large'' gauge transformations is finite or not.
If it is finite and, therefore, physical then we find a net 
topological contribution ${\cal C}$ to $g_A^{(0)}$ \cite{bass99}
\begin{equation}
g_A^{(0)} 
= 
\Biggl(
\sum_q \Delta q - f {\alpha_s \over 2 \pi} \Delta g \Biggr)_{\rm partons}
+ \ {\cal C}
\end{equation}
The topological term ${\cal C}$ has support only at $x=0$.
It is missed by polarised deep inelastic scattering 
experiments which measure $g_1(x,Q^2)$ between some small 
but finite value $x_{\rm min}$ and an upper value $x_{\rm max}$ 
which is close to one.
As we decrease $x_{\rm min} \rightarrow 0$ we measure the first moment
\begin{equation}
\Gamma \equiv \lim_{x_{\rm min} \rightarrow 0} \ 
\int^1_{x_{\rm min}} dx \ g_1 (x,Q^2).
\end{equation}
This means that the singlet axial charge which is extracted
from polarised deep inelastic scattering is 
the combination $g_A^{(0)}|_{\rm pDIS} = (g_A^{(0)} - {\cal C})$.
In contrast, elastic ${\rm Z}^0$ exchange processes such as 
$\nu p$ elastic scattering \cite{kaplan} 
and parity violation in light atoms \cite{parity} 
measure the full $g_A^{(0)}$.
One can, in principle, measure the topology term ${\cal C}$ 
by comparing the flavour-singlet axial charges which are extracted 
from polarised deep inelastic and $\nu p$ elastic scattering experiments.

If some fraction of the spin of the constituent quark is 
carried by 
gluon topology in QCD,
then the constituent quark model predictions for $g_A^{(0)}$ 
are not necessarily in contradiction with the small value of 
$g_A^{(0)}|_{\rm pDIS}$ extracted from deep inelastic scattering experiments.

The presence or absence of topological $x=0$ polarisation is 
intimately related to the dynamics of $U_A(1)$ symmetry
breaking in QCD. 
A simple dynamical mechanism for producing topological $x=0$
polarisation is provided by Crewther's theory of quark-instanton 
interactions \cite{rjc}.
There, any instanton induced suppression of $g_A^{(0)}|_{\rm pDIS}$ 
is compensated by a net transfer of axial charge or ``spin''from 
partons carrying finite momentum fraction $x>0$ to the flavour-singlet 
topological term at $x=0$ \cite{bass99}.

A large positive $\Delta g$ ($\sim +1.5$ at $Q^2=1$GeV$^2$) 
and 
topological $x=0$ polarisation are two possible explanations 
for the small value of  $g_A^{(0)}|_{\rm pDIS}$. 
Measurements of both quantities are urgently needed!

\end{document}